\documentclass[12pt]{article}
\usepackage{amssymb}

 \def \bea { \begin{eqnarray}}
\def \eea {\end{eqnarray}}
\def \be {\begin{equation}}
\def \ee {\end{equation}}

\title{{\bf Gravitational instability of an extreme Kerr black hole}}

\author{James Lucietti$^a$\footnote{j.lucietti@ed.ac.uk } \ and Harvey S. Reall$^b$\footnote{hsr1000@cam.ac.uk} \\ \\
\\ \small \sl $^a$  School of Mathematics and Maxwell Institute of Mathematical Sciences, \\ \small \sl University of Edinburgh, King's Buildings, Edinburgh, EH9 3JZ, UK  \\ \small \sl $^b$ Department of Applied Mathematics and Theoretical Physics, \\ \small \sl Centre for Mathematical Sciences, University of Cambridge, \\ \small \sl Wilberforce Road, Cambridge CB3 0WA, UK} 

\date{}

\begin{document}

\maketitle

\begin{abstract}
Aretakis has proved the existence of an instability of a massless scalar field at the horizon of an extreme Kerr or Reissner-Nordstr\"om black hole: for generic initial data, a transverse derivative of the scalar field at the horizon does not decay, and higher transverse derivatives blow up. We show that a similar instability occurs for linearized gravitational, and electromagnetic, perturbations of an extreme Kerr black hole. We show also that the massless scalar field instability occurs for extreme black hole solutions of a large class of theories in various spacetime dimensions.
\end{abstract}

\section{Introduction}

Extreme (zero temperature) black holes are of special interest because they do not emit Hawking radiation. Hence they are expected to have a simpler description in any candidate theory of quantum gravity. This expectation has been realised within string theory, which has been used to give a statistical mechanics derivation of the Bekenstein-Hawking entropy for certain supersymmetric (therefore extreme) black hole solutions to various supergravity theories \cite{Strominger:1996sh}.  Recently, there has been considerable interest in the proposal that an extreme Kerr black hole can be described by a conformal field theory \cite{Guica:2008mu}.

Given their importance, it is natural to ask: are extreme black hole stable? We will say that an extreme black hole is stable if any perturbation that is small initially remains small for all time and, at late time, ``settles down" to a stationary perturbation corresponding to a small variation of parameters within the family of stationary black hole solutions to which the extreme black hole belongs. (Such a variation of parameters generically makes the black hole slightly non-extreme.)

A heuristic argument suggests that extreme black holes might be classically unstable \cite{Marolf:2010nd}.  Near-extreme black holes usually possess an inner horizon which is believed to be unstable. In the extreme limit, the inner and outer horizons coincide, which suggests that the outer (i.e. event) horizon might be unstable in this limit.\footnote{An extreme rotating black hole also has a quantum mechanical instability involving spontaneous emission of superradiant quanta. We will discuss only classical stability.}

Before discussing the stability of extreme black holes, we will review briefly some stability results for non-extreme black holes. Consider a massless scalar field $\psi$ in the Schwarzschild spacetime. The scalar field can be regarded as a toy model for the more interesting case of linearized gravitational  perturbations. Pick a spacelike hypersurface $\Sigma_0$ which intersects the future horizon ${\cal H}^+$ and extends to null or spacelike infinity. Prescribe initial data for the scalar field on $\Sigma_0$ which vanishes at an appropriate rate at infinity. Let $\Sigma_\tau$ denote the surface obtained by translating $\Sigma_0$ into the future a parameter distance $\tau$ along the orbits of the timelike Killing vector field. It has been proved (see Ref. \cite{Dafermos:2008en} for a review) that the scalar field decays outside ${\cal H}^+$ and also in a neighbourhood of ${\cal H}^+$.  In particular, {\it along the horizon, $\psi$ and all its derivatives decay} at least as fast as certain negative powers of $\tau$. Similar stability results have been achieved for a massless scalar field in a non-extreme Reissner-Nordstr\"om \cite{Blue:2005nj} or non-extreme Kerr \cite{Dafermos:2010hd} spacetime. 

Consider now the case of an extreme black hole. Recently, strong evidence for the existence of a classical {\it instability} has been obtained by Aretakis. He has considered the evolution of a massless scalar field $\psi$ in the background of an extreme Reissner-Nordstr\"om black hole. He proved that, for arbitrary initial data specified on a spacelike surface $\Sigma_0$ intersecting the future even horizon ${\cal H}^+$, $\psi$ decays on and outside ${\cal H}^+$ \cite{Aretakis:2011ha}. However, transverse derivatives of $\psi$ do not decay on ${\cal H}^+$: if $(v,r,\theta,\phi)$ denote advanced Eddington-Finkelstein coordinates then, for generic initial data, $\partial_r \psi$ does not decay on ${\cal H}^+$ and $\partial_r^k \psi$ blows-up as $v^{k-1}$ for large $v$ \cite{Aretakis:2011hc}.  

Aretakis has also investigated the case of a massless scalar field $\psi$ in an extreme Kerr spacetime. He has proved decay of axisymmetric solutions $\psi$, on and outside ${\cal H}^+$ \cite{Aretakis:2011gz}. However, just as in the Reissner-Nordstr\"om case, he finds that, for generic axisymmetric initial data, derivatives of $\psi$ transverse to ${\cal H}^+$ do not decay, and higher order transverse derivatives blow-up along ${\cal H}^+$ \cite{Aretakis:2012ei}. 

In this paper, we will consider linearized {\it gravitational} perturbations of an extreme Kerr black hole. 
Aretakis' results suggest that such perturbations might exhibit instabilities in extreme black hole spacetimes. We will prove in section \ref{sec:teukolsky} that this is indeed the case.  We do this by showing that Aretakis' arguments can be applied to the Teukolsky equation governing linearized gravitational (or electromagnetic) perturbations of Kerr. Our result implies that small linearized gravitational perturbations of an extreme Kerr black hole generically do not settle down to the stationary perturbation corresponding to a small
variation of parameters within the Kerr family of solutions. Hence {\it an extreme Kerr black hole exhibits a linearized gravitational instability}.\footnote{We emphasize that non-extreme Kerr black holes are expected to be stable (at least in vacuum gravity), no matter how small the non-extremality. See ref.~\cite{Dotti:2011ix} for a discussion of the Teukolsky equation {\it inside} a Kerr black hole.}

Section \ref{sec:generalhorizon} presents generalizations of Aretakis' work on massless scalar field instabilities. We prove that his non-decay result can be extended to {\it any} extreme black hole and that his blow-up result extends to extreme black hole solutions of a large class of theories in various dimensions.

\section{Gravitational instability of extreme Kerr}

\label{sec:teukolsky}

\subsection{Naive instability}

Before we introduce our generalisation of Aretakis' work, we will discuss a more obvious candidate instability of an extreme black hole. 

Consider a Kerr-Newman (KN) black hole in Einstein-Maxwell theory. Take an initial spacelike surface $\Sigma_0$ as described above, i.e., intersecting ${\cal H}^+$ and extending to infinity. We assume that $\Sigma_0$ extends a finite distance behind ${\cal H}^+$. Initial data specified on $\Sigma_0$ uniquely determines the black hole solution in the future domain of dependence $D^+(\Sigma_0)$. This region includes those parts of the black hole exterior and event horizon which lie to the future of $\Sigma_0$. If the black hole is non-extreme, it is believed that the solution is nonlinearly stable against arbitrary small perturbations of the initial data on $\Sigma_0$. 

This does not seem to be the case for an extreme black hole. Consider a perturbation of the data on $\Sigma_0$ which corresponds simply to reducing the mass, remaining within the KN family. The effect of this perturbation is drastic: the resulting spacetime is a portion of a super-extreme KN solution, which does not possess an event horizon. 

Is this an instability of the extreme KN solution? To answer this, we must decide what initial data is admissible on a surface such as $\Sigma_0$. In an extreme black hole, $\Sigma_0$ is necessarily geodesically incomplete, terminating either at the singularity or at an inner boundary behind ${\cal H}^+$. Usually one does not consider initial data on such a surface since it is not clear whether the incompleteness is physical. Incompleteness may not be a problem if the singularity is hidden behind a marginally outer trapped surface (MOTS), which is the case when perturbing a non-extreme black hole. But in the extreme case, the perturbed initial data we have just described does not contain a MOTS.

In the non-extreme case, we do not have to confront the problem of dealing with a perturbation specified on an incomplete surface; instead we could choose $\Sigma_0$ to be complete, either extending into a second asymptotically flat region, or intersecting the matter which collapses to form the black hole. But in the extreme case we have no choice: there are no complete spacelike surfaces $\Sigma_0$ which intersect ${\cal H}^+$.\footnote{One might choose $\Sigma_0$ not to intersect the horizon but instead to contain the asymptotic ``throat" region of the extreme black hole geometry. But then we still have the problem of deciding which initial data on $\Sigma_0$ are admissible, i.e., which boundary conditions should be imposed in the throat region.} So how are we to decide which kinds of initial data are admissible on $\Sigma_0$? 

One possibility is to dictate that initial data with an incomplete $\Sigma_0$ is admissible only if the incompleteness is hidden behind a MOTS. Thus extreme KN initial data is admissible but superextreme KN initial data is not. This approach seems unsatisfactory because it simply ``defines away" the possibility of a perturbation destroying the MOTS. 

Alternatively, consider the case in which the extreme black hole forms by gravitational collapse. For example, it is possible to form an extreme Reissner-Nordstr\"om (RN) black hole by spherically symmetric gravitational collapse of charged matter (e.g. see Refs. \cite{shells} for collapse of charged shells). In this case, it is natural to impose initial conditions on a complete asymptotically flat hypersurface that does not intersect the horizon, corresponding to a time before the collapse has occurred. For suitable matter, such initial data will satisfy the mass-charge inequality $M \ge |Q|$ \cite{Gibbons:1982fy}, which excludes the superextreme perturbation just discussed. This supports the view that this perturbation is not admissible for extreme RN. However, it does not appear possible to exclude the superextreme perturbation of extreme Kerr by this kind of argument.\footnote{
One problem is that the (vacuum) mass-angular momentum inequality $M \ge \sqrt{|J|}$ \cite{Dain:2006wb} requires axisymmetry, so this inequality cannot exclude the possibility of a spacetime containing a complete hypersurface on which the initial data is superextreme Kerr outside a compact set, and nonaxisymmetric inside this set.}

To summarise: we have observed that the question of stability of an extreme black hole involves subtleties not present in the non-extreme case. These prevent us from determining the admissibility of the superextreme perturbation of extreme Kerr. Nevertheless, in the next section, we will argue that generic admissible initial data will lead to a gravitational instability of extreme Kerr. 

\subsection{Teukolsky equation for extreme Kerr}

Let $\{ \ell, n, m, \bar{m} \}$ be a null tetrad. Using this we can define the Newman-Penrose Weyl scalars $\Psi_A$, $A=0,1,2,3,4$. A transformation $m \rightarrow e^{i \alpha} m$ is called a {\it spin} and a quantity $\psi$ has {\it spin-weight} $s$ if $\psi \rightarrow e^{i s \alpha} \psi$ under a spin.  For example, $\Psi_A$ has $s=2-A$. The Kerr spacetime is type D, which means that we can choose the tetrad so that only $\Psi_2$ is non-vanishing. Now consider a linearly perturbed Kerr spacetime. Take the tetrad to be an arbitrary linear perturbation of the one just discussed. Then $\delta \Psi_0$ and $\delta \Psi_4$ (the perturbations in $\Psi_0$ and $\Psi_4$) are invariant under infinitesimal diffeomorphisms and infinitesimal changes in the tetrad \cite{Teukolsky:1973ha}. Teukolsky showed that the gauge-invariant quantities $\delta \Psi_0$ and $\delta \Psi_4$ each satisfies a second order wave equation. These two equations take the same form if written in terms of $\delta \Psi_0$ or $\Psi_2^{-4/3} \delta \Psi_4$ respectively \cite{Teukolsky:1973ha}.

Starting from the Kerr metric in Boyer-Lindquist coordinates $(t,r,\theta,\phi)$, convert to Kerr coordinates $(v,r,\theta,\chi)$ defined by
\be
 dv = dt + \frac{r^2 + a^2}{\Delta} dr, \qquad d\chi = d\phi + \frac{a}{\Delta} dr
\ee
where $\Delta = r^2 -2Mr + a^2$ (we will not assume extremality yet). This gives a coordinate chart regular across ${\cal H}^+$, which is at $\Delta=0$. Choose the following tetrad for the background Kerr spacetime:
\bea
 \ell &=& 2 (r^2+a^2) \frac{\partial}{\partial v} + 2a \frac{\partial}{\partial \chi} + \Delta \frac{\partial}{\partial r}, \qquad \qquad n = - \frac{1}{2(r^2 + a^2 \cos^2 \theta)} \frac{\partial}{\partial r}, \nonumber \\ m &=& \frac{1}{\sqrt{2} ( r+ i a \cos \theta)} \left( ia \sin \theta \frac{\partial}{\partial v} + \frac{\partial}{\partial \theta} + \frac{i}{\sin \theta} \frac{\partial}{\partial \chi} \right)  \; .
\eea
The vector fields $\ell$ and $n$ coincide with the principal null directions, with $\ell$ tangential to ${\cal H}^+$. 
This tetrad is regular in a neighbourhood of ${\cal H}^+$ except at $\theta=0,\pi$. By performing a spin one can introduce a new tetrad which is regular at either $\theta=0$ or $\theta=\pi$, but it is not possible to define a tetrad which is globally regular with $\ell,n$ aligned with the principal null directions. Instead one has to work with different tetrads related by spins on coordinate chart overlaps (a spin with $\alpha=\pm \chi$ gives a tetrad regular at $\theta=0,\pi$). This is not a problem because the Teukolsky equation can be written in a form which is manifestly covariant under spins \cite{Stewart:1974uz} although we will not use this form here.

For the above choice of tetrad and coordinates, the Teukolsky equation is\footnote{Note that $\ell=\Delta \ell^K$, $n = \Delta^{-1} n^K$ where a superscript $K$ refers to the Kinnersley tetrad used in Ref. \cite{Teukolsky:1973ha}. This change of tetrad results in a corresponding change in the quantity occurring in the Teukolsky equation: $\psi = \Delta^{s} \psi^K$. So an easy way to obtain the Teukolsky equation in the tetrad and coordinates used here is to take the equation given in Ref. \cite{Teukolsky:1973ha}, substitute $\psi^K = \Delta^{-s} \psi$, multiply by $\Delta^{s}$, and convert to Kerr coordinates.}
\bea
 &&\frac{\partial}{\partial v} \left\{  N(\psi) + 2a \frac{\partial \psi}{\partial \chi}+ 2\left[ (1-2s) r- ias\cos\theta \right] \psi \right\} \nonumber
\\ &&\qquad \qquad = {\cal O} \psi - \Delta \frac{\partial^2 \psi}{\partial r^2} - 2(r-M)(1-s) \frac{\partial \psi}{\partial r} - 2a \frac{\partial^2 \psi}{\partial \chi \partial r} \label{Teukolsky}
\eea
where we have introduced the smooth vector field
\be
\label{N}
 N = 2(r^2+a^2) \frac{\partial}{\partial r} + a^2 \sin^2 \theta \frac{\partial}{\partial v}
\ee
and the operator
\be
 {\cal O} \psi = - \frac{1}{\sin \theta} \frac{\partial}{\partial \theta} \left( \sin \theta \frac{\partial \psi}{\partial \theta} \right) - \frac{1}{\sin^2 \theta} \frac{\partial^2 \psi}{\partial \chi^2} - 2is \frac{\cos \theta}{\sin^2 \theta} \frac{\partial \psi}{\partial \chi} + (s^2 \cot^2 \theta + s ) \psi   \; .
\ee
Note that $N$ is transverse to ${\cal H}^+$. The quantity $\psi$ appearing in the above equation is determined by the value of $s$: $\psi=\delta \Psi_0$ if $s=2$ and $\psi=\Psi_2^{-4/3} \delta \Psi_4$ if $s=-2$. For $s=0$ this equation is just the massless scalar wave equation. Electromagnetic perturbations correspond to $s=\pm 1$ \cite{Teukolsky:1973ha}. 

The operator ${\cal O}$ appears in the theory of spin-weighted spherical harmonics. Using the notation of Ref. \cite{Goldberg:1966uu}, we have ${\cal O} = - \eth\bar{\eth} = - \bar{\eth} \eth + 2s$. It is readily checked that, with respect to the standard measure on the unit sphere $d\Omega=\sin\theta d\theta \wedge d\chi$, the adjoint of $\eth$ is $-\bar{\eth}$ and hence $\mathcal{O}$ is a non-negative self-adjoint operator. The eigenfunctions of ${\cal O}$ are the spin-weighted spherical harmonics ${}_s Y_{jm}$. These are defined for $j=|s|,|s|+1, \ldots$ and $|m| \le j$ with eigenvalues
\be
 {\cal O} ({}_s Y_{jm}) = \left[ j(j+1) - s(s-1) \right] ({}_s Y_{jm}) \; .
\ee
We will assume that $s$ is an integer, hence so is $j$. The eigenspace with zero eigenvalue, which is given by $j=-s$, exists only for $s \leq 0$ and is equal to the kernel of $\bar{\eth}$. Note $\partial_\chi ({}_s Y_{jm})=im ({}_s Y_{jm})$, so ${}_s Y_{j0}$ is independent of the azimuthal angle $\chi$. 

So far, the discussion applies to any Kerr black hole but now we restrict to an extreme Kerr black hole: $M=a>0$, with horizon at $r=a$. Let $H(v)$ denote a $S^2$ cross-section of the future event horizon ${\cal H}^+$ i.e., a surface with $r=a$ and constant $v$. We will now follow reasoning similar to that of Aretakis \cite{Aretakis:2012ei} but with spherical harmonics replaced with spin-weighted harmonics.  

First consider $s \le 0$. Restrict (\ref{Teukolsky}) to $r=a$ and project onto ${}_s Y_{j m}$ with $j=-s$ and $m=0$.  The terms on the RHS give zero contribution, showing that the quantity
\be
\label{I0}
I_0^{(s)}= \int_{H(v)} d\Omega  \left(  {}_{s}Y_{-s \, 0} \right)^*  \left\{ N( \psi ) +2a \left[ (1-2s) - is\cos\theta \right] \psi \right\}  
\ee
is independent of $v$, i.e. it is conserved along $\mathcal{H}^+$. For $s=0$ this agrees with the conserved quantity found by Aretakis~\cite{Aretakis:2012ei}.
Let $\Sigma_0$ be a spacelike surface whose intersection with ${\cal H}^+$ is $H(v_0)$. We are free to specify initial data for $\psi$ on $\Sigma_0$. A generic perturbation will have initial data for which $I_0^{(s)}$ is non-zero. Since $I_0^{(s)}$ is conserved, it remains non-zero for all $v>v_0$. It follows that {\it$\psi$ and the $j=-s$ component of its transverse derivative $N(\psi)$ do not both decay along ${\cal H}^+$ as $v \rightarrow \infty$.} 

One might question the assertion that generic initial data gives non-zero $I_0^{(s)}$. Above we discussed the difficulties associated to defining data on an incomplete surface $\Sigma_0$. Perhaps admissible initial data on $\Sigma_0$ always has vanishing $I_0^{(s)}$. To see why not, consider, for simplicity, the case of a massless scalar in extreme RN, instead of extreme Kerr. The results of Ref. \cite{Aretakis:2012ei} (or section \ref{sec:generalhorizon} of the present paper) show that, for a massless scalar in extreme RN, there is a conserved quantity $I$ exactly analogous to $I_0^{(0)}$. As discussed above, one can form an extreme RN black hole by spherically symmetric gravitational collapse of charged matter. In this case, one can take $\Sigma_0$ to be a {\it complete} surface which intersects ${\cal H}^+$ after the matter has fallen through it and intersects the collapsing matter behind ${\cal H}^+$. Let $\Sigma_*$ be a complete asymptotically flat spacelike surface which does not intersect ${\cal H}^+$, i.e., it corresponds to a time before the black hole has formed. It is uncontroversial that we are free to prescribe arbitrary smooth initial data for $\psi$ on $\Sigma_*$ subject to appropriate boundary conditions at infinity. Cauchy evolution gives a one-to-one correspondence between data on $\Sigma_*$ and data on $\Sigma_0$. Hence we are free to specify arbitrary data on $\Sigma_0$. Such data generically has non-vanishing $I$. This is for extreme RN but there is no reason why extreme Kerr should be any different. Hence generic admissible data has non-vanishing $I_0^{(s)}$.\footnote{We are grateful to M. Dafermos for this argument.}

Now, still with $s \le 0$, act on (\ref{Teukolsky}) with the vector field $N$, set $r=M=a$ and again project onto ${}_s Y_{jm}$ with $j=-s$, $m=0$. This gives
\bea
\partial_v  J^{(s)}_0 =  -2(1-s) \int_{H(v)} d\Omega \; \left( {}_sY_{-s 0} \right)^*N(\psi)  \label{dvJ0}
\eea
where
\bea
 J^{(s)}_0(v) &=& \int_{H(v)}  d\Omega \left({}_s Y_{-s 0} \right)^* \left\{ N(N(\psi)) + 2a \left[ (1-2s) - i s \cos \theta \right] N(\psi) \right. \nonumber \\  &{}& \qquad \qquad \qquad \qquad- \left. a^2 \sin^2 \theta \, {\cal O} \psi + 2 a^2 \left[ 4(1-2s)-(1-s) \sin^2 \theta \right] \psi \right\} \nonumber  \\
 &=&  \int_{H(v)}  d\Omega \left({}_s Y_{-s 0} \right)^* \left\{ N(N(\psi)) + 2a \left[ (1-2s) - i s \cos \theta \right] N(\psi) \right. \nonumber \\  &{}& \left. \qquad \qquad \qquad \qquad+ 2a^2 \left[ 2(3-5s)-(4-3s)\sin^2\theta \right] \psi \right\} \label{J0}
\eea
and the second equality follows from integration by parts and using the identity
\be
{\cal O} \left[ \sin^2 \theta ({}_s Y_{-s 0}) \right]= 2[ 2(s-1)+(3-2s)\sin^2\theta] ({}_s Y_{-s 0}) \; .
\ee
Consider the case in which $\psi \rightarrow 0$ along ${\cal H}^+$ as $v \rightarrow \infty$. Conservation of $I_0^{(s)}$ implies
\be
 \int_{H(v)} d\Omega\;\left(  {}_s Y_{-s 0} \right)^* N(\psi) \rightarrow I_0^{(s)}
\ee
as $v \rightarrow \infty$ and therefore
\be
\partial_v J^{(s)}_0 \rightarrow - 2(1-s) I^{(s)}_0 \; .
\ee
For generic initial data, $I^{(s)}_0 \neq 0$ and hence $J_0^{(s)}$ blows up linearly:
\be
J^{(s)}_0 \sim  - \left( 2(1-s) I^{(s)}_0\right) v  \; .
\ee
Inspecting $J_0^{(s)}$ it follows that, if $\psi \rightarrow 0$ then either $N(\psi)$ or the $j=-s$ component of $N(N( \psi))$ must blow up at least as fast as $v$ as $ v\to \infty$ on ${\cal H}^+$ .   

In summary, we have proved that for an axisymmetric\footnote{Projecting to $m=0$ eigenspaces is equivalent to considering axisymmetric perturbations.} perturbation $\psi$, if $s \le 0$ then $\psi$ and the $j=-s$ component of its transverse derivative $N( \psi)$  cannot both decay along ${\cal H}^+$ as $v \rightarrow \infty$. For $s=0$, it is known that $\psi$ does decay  \cite{Aretakis:2011gz}, and this seems likely also for $s \le 0$ (although proving this would involve a detailed global analysis). In this case, the $j=-s$ component of $N(\psi)$ cannot decay and $N(\psi)$ or the $j=-s$ component of $N(N(\psi))$ must blow up at least as fast as $v$ along ${\cal H}^+$ as $v \rightarrow \infty$. Presumably, one could extend the above analysis to prove that higher derivatives of $\psi$ blow up even faster: for $s=0$ Aretakis states that the $k$th transverse derivative blows up as $v^{k-1}$. 

Following Aretakis, we may go further and derive an infinite set of higher order conserved quantities for any $s$. First differentiate (\ref{Teukolsky}) $p$ times with respect to $r$,\footnote{We could act $p$ times with $N$ rather than with $\partial/\partial r$. This also leads to a conserved quantity but it is harder to write down explicitly.} set $r=M=a$ and project onto ${}_s Y_{jm}$. The result is
\bea
&&  \frac{\partial}{\partial v} \int_{H(v)} d\Omega  \left( {}_s Y_{jm} \right)^* \frac{\partial^p}{\partial r^p} \left\{ N(\psi)+ 2\left[ (1-2s) r- ias\cos\theta + ima \right] \psi \right\} \nonumber \\ && \qquad \qquad = \int_{H(v)} d\Omega  \left( {}_s Y_{jm} \right)^* \left[ (j+p+1-s)(j-p+s) \frac{\partial^p \psi}{\partial r^p} - 2ima\frac{\partial^{p+1} \psi}{\partial r^{p+1}} \right]  \; .
\label{projteuk}
\eea
Note the surprising simplification of the RHS with the first three terms on the RHS of (\ref{Teukolsky}) reducing to a single term.  Now set $m=0$ and $j=p-s$. Since $j \ge |s|$, we must take $p \ge \max(0,2s)$. The RHS above is now zero and hence we have an infinite set of conserved quantities:
\be
I_{p}^{(s)} = \int_{H(v)}  d\Omega \; \left( {}_{s}Y_{p-s \, 0} \right)^*  \;  \frac{\partial^p}{\partial r^p}  \left\{ N(\psi) + 2\left[ (1-2s) r- ias\cos\theta \right] \psi \right\}   \; .
\ee
Note that for $s \leq 0$ we may take $p=0$ which reduces to our earlier conserved quantity (\ref{I0}).  

To obtain higher derivative analogues of $J_0^{(s)}$, use equation (\ref{projteuk}) with $p\to p+1$, $j=p-s$ and $m=0$, which gives
\be
\partial_v J_p^{(s)} = -2(p+1-s) \int_{H(v)} d\Omega \;  ({}_{s}Y_{p-s \, 0} )^*  N\left(\frac{\partial^p \psi}{\partial r^p} \right)
\ee
where
\bea
J_p^{(s)}(v) &=& \int_{H(v)}  d\Omega \; \left( {}_{s}Y_{p-s \, 0} \right)^*  \left( 4a^2\frac{\partial^{p+1}}{\partial r^{p+1}} \left\{ N(\psi)+ 2\left[ (1-2s) r- ias\cos\theta \right] \psi \right\} \right. \nonumber  \\  && \qquad \qquad \qquad \qquad \qquad \left.-     2(p+1-s) a^2 \sin^2\theta \frac{\partial^{p}\psi}{\partial r^{p}}\right) \; .
\eea
Note that for $s \leq 0$ and $p=0$ this again agrees with our earlier formulas (\ref{dvJ0}) and (\ref{J0}).

Now consider $s>0$. The smallest permitted value of $p$ in $I_p^{(s)}$ is $p=2s$ so the argument starts from the conserved quantity $I_{2s}^{(s)}$. Generically this will be non-zero, from which it follows that at least one of the following quantities cannot decay along ${\cal H}^+$: $\partial_r^{2s-1} \psi$, $\partial_r^{2s} \psi$ and the $j=s$ component of $\partial_r^{2s} (N(\psi))$. Hence the best one can hope for is decay of $\psi$ and its first $2s$ derivatives and non-decay of $\partial_r^{2s+1}\psi$. In this case, using $[N, \partial_r^p]= -4p r \partial_r^p-2p(p-1) \partial_r^{p-1}$, implies that $\partial_v J^{(s)}_{2s} \to -2(s+1) I^{(s)}_{2s}$, and hence
\be
J_{2s}^{(s)} \sim   -\left(2(s+1) I^{(s)}_{2s} \right) v
\ee
as $ v\to \infty$, which implies that $\partial_r^{2s+2} \psi$ must blow-up.


Let us now apply these results to linearized gravitational perturbations ($s=\pm 2$). If the extreme Kerr black hole were stable then an arbitrary initial perturbation would settle down to a stationary perturbation corresponding to a small variation of parameters within the Kerr family of solutions. Such a perturbation preserves the type D condition and so has $\delta \Psi_0 = \delta \Psi_4 \equiv 0$. Hence, if the black hole were stable, we could evaluate $I_p^{(s)}$ at large $v$ to deduce $I_p^{(s)} = 0$. It follows that initial data for which one of the $I_p^{(s)} \neq 0$, cannot settle down to such a stationary perturbation and hence {\it the extreme Kerr solution has a linearized gravitational instability.} 

Furthermore, we learn that if $\delta\Psi_4$ decays then a transverse derivative of $\delta \Psi_4$ generically does not decay along ${\cal H}^+$ and certain second transverse derivatives will blow up along ${\cal H}^+$. If $\delta \Psi_0$ and its first $4$ derivatives decay then a $5$th transverse derivative generically will not decay, and a $6$th transverse derivative will blow up. It appears that the Weyl component perturbation $\delta \Psi_4$ exhibits worse behaviour that $\delta \Psi_0$. Note that the former involves 2 factors of the transverse basis vector field $n^a$ in its definition ($\Psi_4 = n^a \bar{m}^b n^c \bar{m}^d C_{abcd}$) whereas the latter involves only tangential basis vector fields ($\Psi_0 = \ell^a m^b \ell^c m^d C_{abcd}$). This means that $\Psi_4$ corresponds to the most tangential components of the Weyl tensor ($C \sim \ell m \ell m$) and $\Psi_0$ to the most transverse ($C \sim n \bar{m} n \bar{m}$). The former is usually associated with outgoing radiation and the latter with ingoing radiation.

It is natural to ask about the evolution of this linearized instability in the full nonlinear theory. One possibility is that a small initial perturbation becomes large but, nevertheless, the spacetime eventually settles down to a slightly non-extreme Kerr black hole. Another possibility is that the spacetime develops a null singularity instead of a horizon \cite{Marolf:2010nd}. 

\section{Scalar field instability of general extreme horizons}

\label{sec:generalhorizon}

In this section, we will extend Aretakis' argument for an instability of a massless scalar field in certain four-dimensional axisymmetric extreme black hole spacetimes~\cite{Aretakis:2012ei}. We will show that his non-decay result can be generalized to {\it any} extreme black hole, and his blow-up result can be generalized to extreme black hole solutions of a large class of theories in various dimensions.

We will work in Gaussian null coordinates \cite{isenberg}, which for convenience we now recall. Let $(M,g)$ be a $D$-dimensional spacetime and $\mathcal{H}^+$ a smooth, degenerate, Killing horizon of a Killing vector field $K$. Let $H_0$ be a $D-2$ dimensional spacelike submanifold of $\mathcal{H}^+$ and assume that each orbit of $K$ is isomorphic to $\mathbb{R}$ and intersects $H_0$ precisely once.\footnote{These assumptions are satisfied by all known stationary extreme black hole solutions.  Such Killing horizons can also arise in non black hole spacetimes.} 
The manifold $H_0$ is called a cross-section and below we will assume these are compact.  The degeneracy condition means that the Killing vector $K$ is tangent to {\it affinely} parameterised null generators of $\mathcal{H}^+$: let $\hat{V} \in \mathbb{R}$ be this affine parameter distance from $H_0$. Let $(\hat{x}^a)$ be coordinates on $H_0$ containing some point $p \in H_0$. This defines coordinates $(\hat{V},\hat{x}^a)$ in a tubular neighbourhood of the integral curve of $K$ through $p$ in $\mathcal{H}^+$ (the $\hat{x}^a$ are extended into this neighbourhood by being taken to be constant along integral curves of $K$).  

Now let $U$ be the unique past-directed null vector field on $\mathcal{H}^+$ satisfying $U \cdot K=1$ and $U \cdot \partial/\partial x^a=0$. Assign coordinates $(\hat{V},\hat{\lambda},\hat{x}^a)$ to the point affine parameter distance $\hat{\lambda}$ along the null geodesic starting at the point on ${\cal H}^+$ with coordinates $(\hat{V},\hat{x}^a)$ with tangent vector $U$ there. These are called Gaussian null coordinates. In these coordinates $K=\partial / \partial \hat{V}$, $U= \partial / \partial \hat{\lambda}$ and it can be shown that the metric is
\be
ds^2 = \hat{\lambda}^2 \hat{F} d\hat{V}^2 +2 d\hat{V}d\hat{\lambda} + 2\hat{\lambda} \hat{h}_a  d\hat{V} d\hat{x}^a +\hat{\gamma}_{ab} d\hat{x}^a d\hat{x}^b
\ee
where all components are smooth functions of $(\hat{\lambda}, \hat{x}^a)$. Degeneracy of the horizon is what implies $g_{\hat{V}\hat{V}}={\cal O}(\hat{\lambda}^2)$. The coordinates $(\hat{x}^a)$ in the above construction are arbitrary; under a change of these coordinates, $\hat{h}_a$ and $\hat{\gamma}_{ab}$ transform as the components of a 1-form and a Riemannian metric on $H_0$.

It is convenient to generalize these coordinates slightly by using a different affine parameter along the geodesics. Define coordinates $(V,\lambda,x^a)$ by $\hat{V}=V$, $\hat{\lambda} = \Gamma(x) \lambda$ and $\hat{x}^a=x^a$ where $\Gamma$ is a smooth positive function. The metric becomes
\be
 ds^2 = \lambda^2 F dV^2 + 2 \Gamma dV d\lambda + 2 \lambda h_a dV dx^a + \gamma_{ab} dx^a dx^b
\ee
where $F=\Gamma^2\hat{F}$, $h_a = \Gamma \hat{h}_a + \partial_a \Gamma$, $\gamma_{ab}=\hat{\gamma}_{ab}$ are all smooth functions of $(\lambda, x^a)$. Let $S(V,\lambda)$ denote a surface of constant $(V,\lambda)$, and $D_a$ the covariant derivative induced on $S(V,\lambda)$. Note that $H(V) \equiv S(V,0)$ is a cross-section of the horizon and $H(0)=H_0$. It turns out that there is a preferred choice for the function $\Gamma$:\footnote{
In all examples known to us, this choice of $\Gamma$ ensures that $h^a$ is a Killing vector field on $H_0$. However, we will not assume this.}\\

\noindent{\bf Lemma 0}. There exists a unique (up to scale), smooth, positive function $\Gamma$ on $H_0$ such that $(D_a h^a)|_{\lambda=0}=0$. \\

\noindent{ \it Proof:} On $H_0$, write $-D_a h^a=- D^2 \Gamma- D_a( \hat{h}^a \Gamma) \equiv \mathcal{L} \Gamma$. We need to show existence of a positive solution of the elliptic partial differential equation ${\cal L}\Gamma=0$. Any 2nd order smooth linear elliptic operator on a compact manifold possesses a principal eigenvalue $\mu$ (which is real and less that or equal to the real part of any other eigenvalue), whose associated eigenfunction $\phi$ is everywhere positive and unique up to scaling~\cite{Andersson:2007fh}. Integrating $\mathcal{L} \phi =\mu \phi$ over $H_0$,  then implies  $\mu \int_{H_0} \phi=0$ and hence, since $\phi>0$ everywhere, $\mu=0$. Therefore $\mathcal{L} \phi=0$ and hence taking $\Gamma$ to be (up to scale) the principal eigenfunction of $\mathcal{L}$ gives the required function. \\

We will consider a massless scalar field $\psi$ in the above geometry. Initial data is prescribed on the spacelike surface $\Sigma_0$ intersecting ${\cal H}^+$ and we assume that boundary conditions are imposed so that $\psi=0$ at infinity. Hence, if stable, $\psi$ should decay along ${\cal H}^+$. 

Writing out the massless scalar wave equation in the above coordinates gives
\bea
0=\Gamma \sqrt{\gamma} \Box \psi &=& \partial_V \left[ \sqrt{\gamma} \left( 2  \partial_\lambda \psi + \frac{\partial_{\lambda} \gamma}{2\gamma} \psi \right) \right] - \partial_{\lambda} \left[ \lambda^2 \sqrt{\gamma} A \partial_{\lambda} \psi \right] - \partial_{\lambda} \left( \lambda \sqrt{\gamma} h^a \partial_a \psi \right) \nonumber \\ &-& \lambda\partial_a \left(\sqrt{\gamma} h^a \partial_{\lambda} \psi \right) + \partial_a \left( \Gamma \sqrt{\gamma} \gamma^{ab} \partial_b \psi \right)
\eea
where $\gamma = \det \gamma_{ab}$, $h^a = \gamma^{ab} h_b$, $\gamma^{ab}$ is the inverse of $\gamma_{ab}$, and we have defined the function
\be
A= \frac{F- h_ah^a}{\Gamma}  \; .
\ee 
Integrate the above equation over $S(V,\lambda)$: the final two terms are total derivatives and so drop out, leaving
\be 
\label{integeq}
  \partial_V \int_{S(V,\lambda)} \sqrt{\gamma} \left( 2  \partial_{\lambda} \psi + \frac{\partial_{\lambda} \gamma}{2\gamma} \psi \right) = \partial_{\lambda} \left\{\lambda^2 \int_{S(V,\lambda)}  \sqrt{\gamma} A\, \partial_{\lambda} \psi - \lambda \int_{S(V,\lambda)}  \sqrt{\gamma} \left( D_a h^a \right) \psi  \right\}
\ee
where in the final term we have integrated by parts. We can now state the first main result of this section:\\ 

\noindent {\bf Lemma 1}.  Choose $\Gamma$ as in Lemma 0. Then the following quantity is a constant along ${\cal H}^+$ (i.e. independent of $V$):
\be
I= \int_{H(V)} \sqrt{\gamma} \left( 2  \frac{\partial \psi}{\partial \lambda} + \frac{\partial_{\lambda} \gamma}{2\gamma} \psi \right) 
\ee

\noindent {\it Proof}. Evaluate (\ref{integeq}) at $\lambda=0$ and use $D_ah^a|_{\lambda=0}=0$. \\

\noindent Note that $\partial_{\lambda} \gamma/(2\gamma) = \Gamma \nabla_\mu (\Gamma^{-1} (\partial/\partial \lambda)^\mu )$, where $\nabla_\mu$ is the spacetime covariant derivative, hence this is a smooth quantity. It is also worth noting that converting back to Gaussian null coordinates gives
\be
I= \int_{H(\hat{V})} \sqrt{\hat{\gamma}}\, \Gamma \left[  2U(\psi) + (\nabla_{\mu} U^\mu) \psi \right]  \; .
\ee
It is easy to see this conserved quantity agrees with that for extreme RN~\cite{Aretakis:2012ei}. We have also checked that it agrees with the conserved quantity (\ref{I0}) (with $s=0$) for extreme Kerr~\cite{Aretakis:2012ei}.\footnote{In particular, we have checked $\Gamma \, U(\psi)|_{\hat{\lambda}=0}= (4a^2)^{-1}N(\psi)|_{r=a}$ for any axisymmetric function $\psi$, where $N$ is the vector field (\ref{N}), and $\Gamma=(1+\cos^2\theta)/2$ can be read off from the near-horizon geometry (see e.g.~\cite{Kunduri:2008rs}).  It then easily follows that  $I=I_0^{(0)}$, where $I^{(0)}_0$ is the conserved quantity (\ref{I0}) with $s=0$.}\\

\noindent {\bf Corollary 1}. Generic initial data has $I \ne 0$ and hence, for such data, $\psi$ and $\partial_{\lambda} \psi$ cannot both decay along ${\cal H}^+$ as $v \rightarrow \infty$.  \\

\noindent This is a non-decay result that applies to {\it any} extreme black hole. To demonstrate blow-up we need an extra assumption about the black hole, whose validity we will discuss at the end of this section.\\

\noindent{\bf Lemma 2}.  Let $\Gamma$ be a smooth positive function on $H_0$ as in Lemma 0. Suppose further that $A|_{\lambda=0}=A_0$ where $A_0 \neq 0$ is a constant. Let
\be
 J(V) \equiv  \int_{H(V)} \partial_{\lambda} \left[  \sqrt{\gamma} \left( 2 \partial_\lambda \psi + \frac{\partial_{\lambda} \gamma}{2\gamma} \psi \right) \right] 
 \ee 
If $\psi \to 0$ along $\mathcal{H}^+$ as $V \to \infty$ and $I \neq 0$, then $J(V)$ blows up linearly: $J(V) \sim A_0 I V$ along $\mathcal{H}^+$ as $V \to \infty$. \\
  
\noindent {\bf Proof}.  Act on (\ref{integeq}) with $\partial_{\lambda}$ and evaluate at $\lambda=0$ to obtain
\be
\label{integeq2}
 \partial_V J(V) = 2 \int_{H(V)}  \sqrt{\gamma} \left[ A \partial_{\lambda} \psi - \partial_{\lambda} \left( D_a h^a \right) \psi  \right]
\ee 
By assumption $\psi \rightarrow 0$, so the final term on the RHS of (\ref{integeq2}) decays and the first term on the RHS asymptotically approaches $A_0 I$. Therefore as $V \to \infty$, $\partial_V J(V) \rightarrow A_0 I$ and integrating this  proves the result. \\

\noindent {\bf Corollary 2}. If $\psi \to 0$ along $\mathcal{H}^+$ as $ V\to \infty$ for generic initial data then either $\partial_{\lambda} \psi$ or $\partial_{\lambda}^2 \psi$ diverges along ${\cal H}^+$ as $V \rightarrow \infty$ (and if $\partial_{\lambda} \psi$ diverges then it most do so consistently with constancy of $I$). \\

In summary, we have shown that, for generic initial data we must have one of the following possibilities: (i) $\psi$ does not decay along ${\cal H}^+$, or (ii) $\psi$ decays, $\partial_{\lambda} \psi$, does not decay and, subject to the assumption about $A$ of Lemma 2, one of the quantities $\partial_{\lambda} \psi$, $\partial_{\lambda}^2 \psi$ blows up as $V \rightarrow \infty$ along ${\cal H}^+$. The ``most stable" outcome consistent with our results is (ii) with $\partial_{\lambda} \psi$ non-decaying but bounded and $\partial_{\lambda}^2 \psi$ blowing up.  \\

Let us return to the assumption in Lemma 2. Since this involves a quantity {\it intrinsic} to the horizon, it can be regarded as an assumption about the near-horizon geometry of the extreme black hole in question (defined by $V \rightarrow V/\epsilon$, $\lambda \rightarrow \epsilon \lambda$ and $\epsilon \rightarrow 0$, see e.g.~\cite{Kunduri:2007vf}). 

This assumption is true for a large class of near-horizon geometries in various dimensions and theories. All extreme black holes solutions known to us satisfy this assumption. For many examples, it follows from the near-horizon $AdS_2$-symmetry theorems proved in Refs. ~\cite{Kunduri:2007vf,Figueras:2008qh}. 
The results of Ref.~\cite{Kunduri:2007vf} imply that the assumption is valid (with $A_0<0$) for extreme black hole solutions of a class of theories in $D=4,5$ dimensions consisting of Einstein gravity coupled to arbitrarily many abelian vectors and uncharged scalars, assuming that the black hole has $D-3$ commuting rotational symmetries and that the horizon topology is non-toroidal. Ref.~\cite{Figueras:2008qh} determined the near-horizon geometries of extreme Myers-Perry black holes \cite{Myers:1986un}, and these also satisfy our assumption with $A_0<0$. In that work, it was also shown that the assumption is valid for $D>5$ extreme vacuum black holes with cohomogeneity-1 near-horizon geometries possessing certain non-abelian rotational symmetry groups. \\

\vspace{.5cm}
\noindent{{\bf Acknowledgements}}:  
We are grateful to Stefanos Aretakis, Piotr Chrusciel, Bob Wald and especially Mihalis Dafermos for useful discussions. JL is supported by an EPSRC Career Acceleration Fellowship. HSR is supported by a Royal Society University Research Fellowship and by European Research Council grant no. ERC-2011-StG 279363-HiDGR. This work was completed while HSR was a participant in the Mathematical Relativity workshop at the Mathematics Institute, Oberwolfach; he is grateful to the workshop organizers and the Institute for hospitality.


\begin{thebibliography}{99}

\bibitem{Strominger:1996sh} 
  A.~Strominger and C.~Vafa,
  Phys.\ Lett.\ B {\bf 379}, 99 (1996)
  [hep-th/9601029].

\bibitem{Guica:2008mu} 
  M.~Guica, T.~Hartman, W.~Song and A.~Strominger,
  Phys.\ Rev.\ D {\bf 80}, 124008 (2009)
  [arXiv:0809.4266 [hep-th]].

\bibitem{Marolf:2010nd} 
  D.~Marolf,
  Gen.\ Rel.\ Grav.\  {\bf 42}, 2337 (2010)
  [arXiv:1005.2999 [gr-qc]].
  
\bibitem{Dafermos:2008en} 
  M.~Dafermos and I.~Rodnianski,
  arXiv:0811.0354 [gr-qc].
  
\bibitem{Blue:2005nj} 
  P.~Blue and A.~Soffer,
  J.\ Funct.\ Anal.\  {\bf 256}, 1 (2009)
  [math/0511281 [math.AP]].
  
\bibitem{Dafermos:2010hd} 
  M.~Dafermos and I.~Rodnianski,
  Proceedings of the 12th Marcel Grossmann meeting, Eds. T. Damour, R.T. Jantzen and R. Ruffini, World Scientific (2012). [arXiv:1010.5137].
  
\bibitem{Aretakis:2011ha} 
  S.~Aretakis,
  Commun.\ Math.\ Phys.\  {\bf 307}, 17 (2011)
  [arXiv:1110.2007 [gr-qc]].

\bibitem{Aretakis:2011hc} 
  S.~Aretakis,
  Annales Henri Poincare {\bf 12}, 1491 (2011)
  [arXiv:1110.2009 [gr-qc]].
     
\bibitem{Aretakis:2011gz} 
  S.~Aretakis,
  arXiv:1110.2006 [gr-qc].

\bibitem{Aretakis:2012ei} 
  S.~Aretakis,
  arXiv:1206.6598 [gr-qc].
   
\bibitem{Dotti:2011ix}
  G.~Dotti, R.~J.~Gleiser and I.~F.~Ranea-Sandoval,
  Int.\ J.\ Mod.\ Phys.\ Proc.\ Suppl.\ E {\bf 20} (2011) 27
  [arXiv:1111.5974 [gr-qc]].
  
 \bibitem{shells}
K. Kuchar, Czech. J. Phys. B{\bf 18}, 435 (1968); Ch. J. Farrugia and P. Hajicek, Comm. Math. Phys. {\bf 68}, 291 (1979).
  
\bibitem{Gibbons:1982fy} 
  G.~W.~Gibbons and C.~M.~Hull,
  Phys.\ Lett.\ B {\bf 109}, 190 (1982).
  
  \bibitem{Dain:2006wb} 
  S.~Dain,
  J.\ Diff.\ Geom.\  {\bf 79}, 33 (2008)
  [gr-qc/0606105].

  
\bibitem{Teukolsky:1973ha} 
  S.~A.~Teukolsky,
  Phys.\ Rev.\ Lett.\  {\bf 29}, 1114 (1972);
  Astrophys.\ J.\  {\bf 185}, 635 (1973).

\bibitem{Stewart:1974uz} 
  J.~M.~Stewart and M.~Walker,
  Proc.\ Roy.\ Soc.\ Lond.\ A {\bf 341}, 49 (1974).

\bibitem{Goldberg:1966uu} 
  J.~N.~Goldberg, A.~J.~MacFarlane, E.~T.~Newman, F.~Rohrlich and E.~C.~G.~Sudarshan,
  J.\ Math.\ Phys.\  {\bf 8}, 2155 (1967).

\bibitem{isenberg}
V. Moncrief and J. Isenberg, Commun. Math. Phys. {\bf 89}, 387 (1983). 

\bibitem{Andersson:2007fh}
  L.~Andersson, M.~Mars and W.~Simon,
  arXiv:0704.2889 [gr-qc].

\bibitem{Kunduri:2008rs}
  H.~K.~Kunduri and J.~Lucietti,
  J.\ Math.\ Phys.\  {\bf 50} (2009) 082502
  [arXiv:0806.2051 [hep-th]].

\bibitem{Kunduri:2007vf} 
  H.~K.~Kunduri, J.~Lucietti and H.~S.~Reall,
  Class.\ Quant.\ Grav.\  {\bf 24}, 4169 (2007)
  [arXiv:0705.4214 [hep-th]].
  
  
  \bibitem{Figueras:2008qh}
  P.~Figueras, H.~K.~Kunduri, J.~Lucietti and M.~Rangamani,
  Phys.\ Rev.\ D {\bf 78} (2008) 044042
  [arXiv:0803.2998 [hep-th]].

\bibitem{Myers:1986un} 
  R.~C.~Myers and M.~J.~Perry,
  Annals Phys.\  {\bf 172}, 304 (1986).

  

\end{thebibliography}
\end{document}